\documentclass[letterpaper, 10 pt, conference]{ieeeconf}  

\IEEEoverridecommandlockouts                              

\usepackage{amsfonts}
\usepackage{amsmath}
\usepackage{algorithmic}
\usepackage{graphicx}
\usepackage{subcaption}
\usepackage{textcomp}
\usepackage{array}
\usepackage{xcolor}
\usepackage{layouts}
\usepackage{tabularx}
\usepackage{booktabs}
\usepackage{color,soul}
\usepackage[nolist]{acronym}
\usepackage{orcidlink}

\newcommand\copyrighttext{%
    \footnotesize \copyright{ }2024 IEEE. Personal use of this material is permitted. Permission from IEEE must be obtained for all other uses, in any current or future media, including reprinting/republishing this material for advertising or promotional purposes, creating new collective works, for resale or redistribution to servers or lists, or reuse of any copyrighted component of this work in other works.}
\newcommand\copyrightnotice{%
    \begin{tikzpicture}[remember picture,overlay]
    \node[anchor=south,yshift=15pt,xshift=0pt] at (current page.south) {\parbox{\dimexpr\textwidth-\fboxsep-\fboxrule\relax}{\copyrighttext}};
    \end{tikzpicture}%
}

\def\BibTeX{{\rm B\kern-.05em{\sc i\kern-.025em b}\kern-.08em
    T\kern-.1667em\lower.7ex\hbox{E}\kern-.125emX}}

\title{\LARGE \bf
Determining the Tactical Challenge of Scenarios to Efficiently Test Automated Driving Systems
}

\author{Lennart Vater$^{1}$, Sven Tarlowski$^{1}$, Michael Schuldes$^{1}$ and Lutz Eckstein$^{1}$
\thanks{The research leading to these results has received funding from the European Union's Horizon 2020 research and innovation programme under grant agreement No 101006664. 
The sole responsibility of this publication lies with the authors. 
The authors would like to thank all partners within the Hi-Drive project (hi-drive.eu) for their cooperation and valuable contribution.}
\thanks{$^{1}$All authors are with the Institute for Automotive Engineering,
        RWTH Aachen University, 52074 Aachen, Germany
        {\tt\small \{vater, tarlowski, schuldes, eckstein\}@ika.rwth-aachen.de}}%
}

\begin{document}

\maketitle
\thispagestyle{empty}
\pagestyle{empty}
\copyrightnotice

\begin{abstract}
The selection of relevant test scenarios for the scenario-based testing and safety validation of automated driving systems (ADSs) remains challenging.
An important aspect of the relevance of a scenario is the challenge it poses for an ADS.
Existing methods for calculating the challenge of a scenario aim to express the challenge in terms of a metric value.
Metric values are useful to select the least or most challenging scenario.
However, they fail to provide human-interpretable information on the cause of the challenge which is critical information for the efficient selection of relevant test scenarios.
Therefore, this paper presents the \emph{Challenge Description Method} that mitigates this issue by analyzing scenarios and providing a description of their challenge in terms of the minimum required lane changes and their difficulty.
Applying the method to different highway scenarios showed that it is capable of analyzing complex scenarios and providing easy-to-understand descriptions that can be used to select relevant test scenarios. 

\end{abstract}

\bstctlcite{IEEEexample:BSTcontrol}

\section{INTRODUCTION} \label{sec:introduction}

Before vehicles equipped with automated driving systems (ADSs)~\cite{j3016} can be deployed, their safety has to be validated in tests.
Current research focuses on the scenario-based approach~\cite{RIE20} that is designed around a scenario database to shorten expensive testing time.
The database is populated with scenarios from real traffic and from these scenarios only the relevant ones will be selected and used as test scenarios~\cite{iso34501} for the ADSs.
However, the selection of the relevant test scenarios remains difficult because the database contains large amounts of scenarios and not all scenarios are relevant for each ADS.
Thus, additional information about the scenarios in the database is needed to efficiently select relevant test scenarios for an ADS.
Expensive testing could be reduced if the database would provide information on how hard it is in general to pass each of the scenarios, i.\,e., how objectively challenging the scenarios are.

A current method for determining the challenge of a scenario is to apply safety metrics like the time-to-collision metric to each time step of the scenario and then aggregate the individual metric values to scenario level~\cite{WES23}.
Another approach is to use the reachability analysis~\cite{SOE17} to determine the theoretically drivable area and use its size as a surrogate for the challenge of the scenario~\cite{WU22}.
However, these approaches either require a high manual calibration effort or fail to provide human-interpretable challenge information that is representative for multiple similar ADSs.

\begin{figure}[t] 
  \includegraphics[width=\columnwidth]{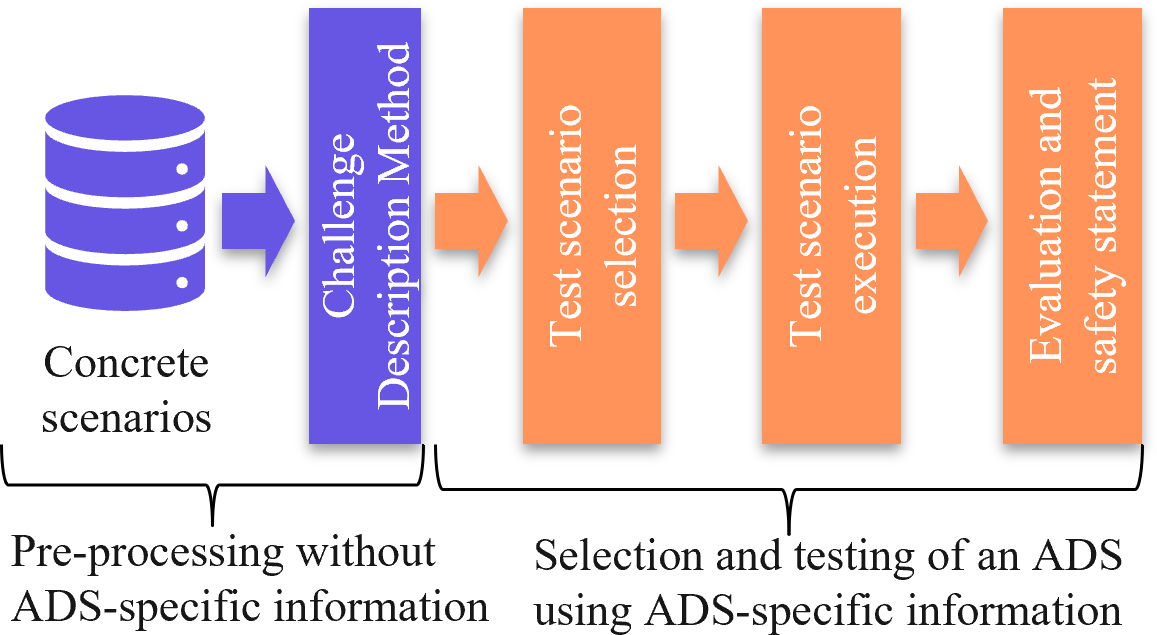}
  \caption{Integration of the proposed Challenge Description Method into the scenario-based testing and safety validation approach. The pre-processing step (blue) is performed only once for multiple similar ADS and the selection and testing steps (orange) once per ADS.}
  \label{fig:context}
\end{figure}

Thus, this paper presents a novel method called \emph{Challenge Description Method} to mitigate the issue.
The method uses the reachability analysis to determine the theoretically drivable area but instead of using the size of the area, it derives the easiest tactical maneuver that is needed to pass the scenario.
The tactical maneuver is given in terms of the minimum number of lane changes required and their difficulty.
Since the method is based on the reachability analysis instead of the safety metric, it does not require complicated calibration but only some simple parameter values need to be set based on the expected capabilities of the ADS.
To prove the usefulness and interpretability of the method, we applied it to four exemplary highway scenarios that differ in terms of contained road users and behavior and present the findings in this paper.

Figure~\ref{fig:context} shows how the proposed method contributes to the scenario-based approach.
In the offline pre-processing step, the method analyzes all scenarios in the database and provides human-interpretable challenge descriptions.
ADS developers then use these descriptions together with other scenario parameters (e.g., the road characteristics) to efficiently filter and select relevant test scenarios for their ADS.

The paper is structured as follows: Section~\ref{sec:related_work} presents the related work, Section~\ref{sec:background} introduces relevant background, and Section~\ref{sec:method} details the actual method.
In Section~\ref{sec:evaluation} we showcase the method by applying it to four exemplary highway scenarios.
The paper is concluded by Section~\ref{sec:conclusion}.

\section{RELATED WORK}\label{sec:related_work}

Most methods that determine the challenge of scenarios employ simple safety metrics~\cite{WES23}, e.g. the time-to-react (TTR) metrics.
However, using these metrics presents the following two challenges: First, most metrics are designed to work only in a single time step, but scenarios contain multiple time steps called \emph{scenes}~\cite{ULB15}.
Therefore, the metric values for each scene must be aggregated to the scenario level.
Second, the scope of a single metric is limited to certain situations, e.g., car-following situations.
Therefore, multiple metrics must be combined which raises the problem of selecting appropriate weights for each metric.

For example, Ponn et al.~\cite{PON20} propose a linear combination of 13 metrics to determine challenging scenarios and conduct an expert study to find the weights.
Their method calculates the metrics per scene and aggregates the score to the scenario level by selecting the maximum value of all scenes.
Nalic et al.~\cite{NAL22} present a different approach that uses a combination of TTR metrics.
Since the scope of these metrics is limited to car-following situations, they introduce \emph{fictive vehicles} that are copies of the SUT in its adjacent lanes to also apply the metrics there.
However, they fail to address the aggregation to the scenario level.

Junietz et al.~\cite{JUN18} presented another approach to determine the challenge of scenarios.
Per scene of a scenario, they used an offline trajectory planner that determines the least difficult to drive trajectory for the system under test (SUT) by minimizing a cost function consisting of a combination of metrics.
The minimum cost function value is the surrogate for the risk in the scene.
They propose to aggregate to the scenario level by means of a maximum or mean.

The listed methods work on scenarios that include the entire behavior of the SUT as well as the surrounding vehicles.
However, when it comes to test scenarios (see Section~\ref{subsec:scenarios}), only the initial position of the SUT is known.
To calculate safety metrics like the TTR metrics for an entire scenario, the position of the SUT must be known in each scene.
Therefore, the methods fail to determine the challenge of test scenarios.

To mitigate the problems of manual selection and calibration of metrics and the prediction of the SUT's behavior, other research focuses on the reachability analysis~\cite{SOE17}.
The reachability analysis can be used to determine the theoretically drivable area of the SUT in a scenario.
For example, Wu et al.~\cite{WU22} proposed to use the \emph{Discretized Normalized Drivable Area} (DNDA) as a surrogate for the challenge of a scenario.
The DNDA is computed in each scene by dividing the size of the drivable area by the size of the drivable area without any surrounding vehicles.
However, they only provide a metric value per scene and do not address the aggregation to the scenario level.

\section{BACKGROUND AND DEFINITIONS} \label{sec:background}

\subsection{Test Scenarios} \label{subsec:scenarios}

For the terms \emph{scene} and \emph{scenario}, we use the definitions of Ulbrich et al.~\cite{ULB15}: A scenario consists of a sequence of scenes and each scene is a snapshot of the environment at a time step.
A concrete scenario, in contrast to a functional or logical scenario, is a scenario that has a fixed content defined by a set of parameter values~\cite{MEN18}.
We will use the term \emph{scenario} instead of \emph{concrete scenario} in the remainder of this paper.

This paper focuses on tactical challenges of scenarios on highway-like roads arising from layer 1 (road) and layer 4 (dynamic objects) of the 6-Layer Model~\cite{SCHO21} and assumes a perfect environment perception.
Perfect environment perception means that the trajectories of all vehicles in the scenario are known and there is no need to predict them.
This assumption is based on the idea that in scenario-based testing, each scenario reflects a certain behavior of the surrounding road users. 
Therefore, a closed-loop simulation is not required to cover all behavior.

We denote a scenario with $S$ and define it as a set of parameter values from these two layers:
\begin{equation}\label{eq:scenario}
    S := \{ L_1, L_4 \}.
\end{equation}
$L_1$ is the set of parameters describing the road network, here, they are the road width and the number of lanes.
$L_4$ is a set consisting of sets of parameters for each road user in the scenario.
In this paper, we consider the positions, velocities, accelerations and orientations over the course of the entire scenario as parameters.

The term \emph{test scenario} refers to a scenario that is used for testing an ADS as defined in ISO 34501~\cite{iso34501}.
When testing an ADS, it becomes the system under test (SUT).
To use a concrete scenario as a test scenario for an ADS, two more components must be added: the initial conditions of the SUT in the scenario and one or more \emph{success criteria}.
Since the test scenario is used to test the SUT, only its initial condition is known prior to the test.
The success criteria are used after the execution of a test scenario to determine, whether the SUT has passed the test scenario.
We define a test scenario $T$ as a set of a scenario $S$, the initial conditions for the SUT $I_0$ and a set $C$ of one or more success criteria:
\begin{equation}\label{eq:test_scenario}
    T := \left\{S, I_0, C \right\}.
\end{equation}

\subsection{Tactical Challenge of Scenarios}\label{subsec:challenge}

Based on the definition of Ponn et al.~\cite{PON20}, we use the term \emph{challenging} as an objective property of a test scenario itself.
We explicitly distinguish challenging from \emph{critical} which is a property of the test scenario together with the capabilities of the SUT.
Furthermore, we define the tactical challenge in a test scenario in relation to the \emph{normal operation} described in the SAE J3016~\cite{j3016}.
We consider ADS of at least SAE Level 4~\cite{j3016}, i.\,e., ADS that perform the complete dynamic driving task.
According to the SAE J3016, normal operation of an ADS can be divided into three functional components: strategic, tactical and operational.
The strategic component involves decisions on the highest level (e.g., the selection of the destination), 
the operational component is concerned with the lowest level of control (e.g., lateral control for lane keeping), and the tactical component is situated in-between the other two.
The tactical component is responsible for monitoring the environment, planning maneuvers and enhancing conspicuity.

This paper focuses on the \emph{maneuver planing} aspect of the tactical component, i.\,e., what tactical maneuvers are required to keep driving in normal operation.
The paper specifically focuses on lane changes as tactical maneuvers.
Therefore, we define the tactical challenge of a scenario by how hard it is for a SUT to stay in normal operation in that scenario, i.\,e. it does not have to perform a minimal risk maneuver.
However, whether a SUT stays in normal operation in a scenario must not necessarily be a success criterion of a test scenario (see Section~\ref{subsec:scenarios}).
On the one hand, in a scenario without any available tactical maneuvers, the ADS cannot stay in normal operation and successfully performing a minimal risk maneuver is a more reasonable success criterion.
On the other hand, ADS developers might define stricter success criteria for their ADS.

\subsection{Reachability Analysis} \label{subsec:reach_set}

The reachability analysis is concerned with computing the reachable set.
The reachable set is the set of states a vehicle can reach after a certain period of time without colliding with other vehicles, going off-road, or exceeding set limits on the velocity or acceleration~\cite{SOE17}.
For the reachable set, we use the definitions and notations introduced by Liu et al.~\cite{LIU22} and S\"ontges et al.~\cite{SOE17}.
Their most relevant definitions for our work are introduced in the following.
For more details, please refer to these two aforementioned publications.

The state of the ego vehicle at a discrete time step $k \in \mathbb{N}_0$ is given by $\pmb{x}_k = (p_{s,k}, v_{s,k}, p_{t,k}, v_{t,k})^\top \in \mathbb{R}^4$.
The subscripts $s$ and $t$ denote the longitudinal and lateral dimension of the position $p$ and velocity $v$ in the curvilinear coordinate frame, respectively.
A state-space discrete-time point-mass model describes the ego vehicle motion with a constant time $\Delta t$ between two discrete time steps $k$. 
The input at each time step $k$ are the acceleration in longitudinal and lateral dimensions, denoted by $\pmb{u}_k = (a_{s,k}, a_{t,k})^\top \in \mathbb{R}^2$.
The vehicle motion is constrained by constant and real-valued minimum and maximum values for the velocities and acceleration:
\begin{equation}\label{eq:paramters}
\begin{split}
    v_{d,min} &\leq v_{d,k} \leq v_{d,max} \\
    a_{d,min} &\leq a_{d,k} \leq a_{d,max}\ \text{with}\ d \in \{ s,t \}.
\end{split}
\end{equation}
As Liu et al.~\cite{LIU22} discussed, using the simple point-mass model over more complex and realistic models for the ego vehicle motion results in an overapproximation of the reachable set.
Thus, the reachable set calculated using the point-mass model includes all reachable states for a real vehicle and some states that cannot be reached by a real vehicle.
As a result, it represents an upper bound for what is reachable.
The states not included in the reachable state cannot be reached by a real vehicle.

The reachable set $\mathcal{R}^*_k$ at a time step $k$ is given by~\cite{LIU22}:
\begin{equation}
\begin{split}
    \mathcal{R}^*_k := \Bigl\{ &\mathcal{X}_k(\pmb{x}_0, \pmb{u}_{[0,k]}) ~|~ \exists \pmb{x}_0 \in \mathcal{X}_0, \forall \tau \in \{0, ..., k \}, \\
    &\exists \pmb{u}_\tau \in \mathcal{U}_\tau : \mathcal{X}_\tau (\pmb{x}_0, u_{[0,\tau]}) \not\in \mathcal{F}_\tau \Bigl\} .
\end{split}
\end{equation}
$\mathcal{X}_k(\pmb{x}_0, u_{[0,k]})$ represents the solution of the state-space formula at time step $k$ when starting at state $\pmb{x}_0$ and taking the input trajectory $\pmb{u}_{[0,k]}$.
$\mathcal{F}_k$ is the set of forbidden states, which are defined by the positions of the object vehicles as well as the area the ego vehicle is occupying at time step $k$.

Since $\mathcal{R}^*_k$ is hard to calculate efficiently~\cite{LIU22}, Liu et al. propose to use an overapproximation $\mathcal{R}_k = \{ \mathcal{R}^{(0)}_k, ..., \mathcal{R}^{(n)}_k \}$ instead.
This overapproximation of the reachable set is the union of $n$ \emph{base sets} $\mathcal{R}^{(i)}_k$ with $i \in \{0, ..., n\}$.
A base set is a Cartesian products of the two polytopes describing the reachable positions and velocities in $s$ and $t$ dimension, respectively.
A reachable set $\mathcal{R}_k$ can be projected to the position domain, yielding the drivable area $\mathcal{D}_k$, which is a union of $\mathcal{D}^{(i)}_k$ with $i \in \{0, ..., n\}$.
A drivable area can be interpreted as the space on the road that can be reached by the vehicle.

To describe the temporal relation between the base sets of multiple time steps Liu et al.~\cite{LIU22} introduce the \emph{reachability graph} $\mathcal{G_R}$.
The reachability graph is a directed graph.
The nodes are the base sets of all time steps $\mathcal{N}(\mathcal{G_R}) = \{\mathcal{R}^{(0)}_0, ..., \mathcal{R}^{(n)}_K \}$ with $K$ being the last time step.
An edge $(\mathcal{R}^{(j)}_{k}, \mathcal{R}^{(j)}_{k+1})$ from a base set $\mathcal{R}^{(i)}_{k}$ in time step $k$ to a base set $\mathcal{R}^{(j)}_{k+1}$ exists if $\mathcal{R}^{(j)}_{k+1}$ can be reached from $\mathcal{R}^{(i)}_{k}$.

\section{METHOD} \label{sec:method}

\begin{figure*}[t] 
  \includegraphics[width=\textwidth]{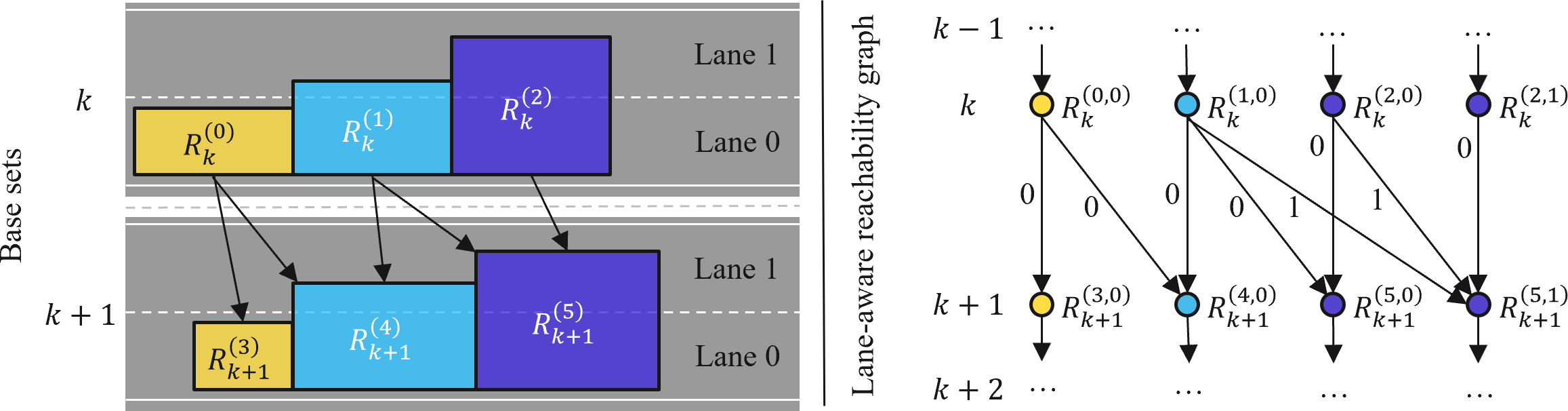}
  \caption{Visualization of the lane-aware reachability graph (right) for base sets projected on the position domain (left).}
  \label{fig:graph}
\end{figure*}

\subsection{Design}\label{subsec:design}

The Challenge Description Method analyzes a scenario and provides a description of the objective tactical challenge.
As defined in Section~\ref{subsec:challenge} the tactical challenge of a scenario means how hard it is for a SUT to stay in normal operation in the scenario.
We define the normal operation by boundaries for the velocity and the acceleration of the SUT, i.\,e., minimum and maximum thresholds in s- and t-direction.
These are the only eight parameters of the method (cf.~Table~\ref{tab:parameters}).
Since we focus on the maneuver planing aspect of the tactical aspect of an ADS and on lane changes as tactical maneuvers, the tactical challenge of a scenario can be one of the following:
\begin{enumerate}
    \item In this scenario, the SUT can stay in normal operation without performing a lane change. It may have to adjust its speed.
    \item In this scenario, the SUT can stay in normal operation by performing one or more lane changes. It may have to adjust its speed.
    \item In this scenario, the SUT cannot stay in normal operation, i.\,e., a minimal risk maneuver is required.
\end{enumerate}

For case 2), the method provides additional details on the tactical challenge: The number of required lane changes and the difficulty of each of the lane changes in form of the decision time window.
The following subsections detail the method's steps.

\subsection{Reachability Analysis}

The method uses the reachability analysis (Section~\ref{subsec:reach_set}) to compute the reachable set for the SUT.
The computation is done using the CommonRoad Reach implementation provided by Liu et al.~\cite{LIU22}.
To use the implementation, the components of the test scenarios from Eq.~\ref{eq:test_scenario} must be mapped to those of CommonRoad motion planning problems~\cite{ALT17}.
The parameters of the scenario $S$ from Eq.~\ref{eq:scenario}, i.\,e., the road parameters $L_1$ and the road user parameters $L_4$, are modelled as \emph{lanelets} and \emph{dynamicObstacles}, respectively.
The initial conditions of the SUT $I_0$ are modelled by the \emph{initialState} in the \emph{planningProblem} attribute of the CommonRoad format.
The success criterion is realized by a \emph{goalState} in the \emph{planningProblem}.

The boundaries for velocity and acceleration defined for the normal operation are used as parameter values to compute the reachable set (see Eq.~\ref{eq:paramters}).
Therefore, the computed reachable set consists of base sets for the normal operation.
The duration of the scenarios is modelled by a goal region.
The method uses the reachable set to determine whether the goal region can be reached.
If there exists no base set with a drivable area that intersects with the goal region, the SUT cannot remain in normal operation in the scenario.
Thus, a minimal risk maneuver is needed in the scenario and the challenge of the scenario corresponds to the case 3).

\subsection{Lane-aware Reachability Graph}\label{subsec:larg}

If the goal region can be reached, the challenge of the scenario is either case 1) or 2) and the required tactical maneuvers, i.\,e., lane changes, must be determined.
To derive lane changes from the reachable set, the temporal connections between the base sets are used.
Therefore, the reachability graph~$\mathcal{G_R}$ (see Section~\ref{subsec:reach_set}) is constructed from the reachable set.
Figure~\ref{fig:graph} shows an example for three base sets in two consecutive time steps on the left side.
The arrows between the upper base sets to the lower base sets indicate which base sets in $k+1$ are reachable from which base sets in $k$.
Thus, they represent the edges in $\mathcal{G_R}$.
A path in the graph corresponds to a sequence of possible states of the SUT.
If projected to the position domain, the graph can be used to determine possible paths the SUT could take on the road.
However, the reachability graph does not include information about the lanes that the base sets are occupying in the position domain.

Therefore, an enhanced version of $\mathcal{G_R}$, the lane-aware reachability graph~$\mathcal{G_{R,L}}$, is constructed.
Each base set $\mathcal{R}^{(i)}_k$ is analyzed to determine which lanes it occupies in the position domain.
The $\mathcal{G_{R,L}}$ that corresponds to the situation on the left side in Figure~\ref{fig:graph} is depicted on the right side in the same figure.
For each lane $j$ that a base set $\mathcal{R}^{(i)}_k$ occupies, a new node $\mathcal{R}^{(i , j)}_k$ is created in $\mathcal{G_{R,L}}$.
A base set occupies a lane if the overlap of its drivable area and the lane's polygon in t-direction is at least the width of the SUT.
In general, $\mathcal{G_{R,L}}$ has at least as many nodes as $\mathcal{G_{R}}$.
In the example, all base sets occupy Lane~0 and $\mathcal{R}^{(2)}_{k}$ and $\mathcal{R}^{(5)}_{k+1}$ additionally occupy Lane~1.
Thus, $\mathcal{G_{R,L}}$ consists of eight nodes for the six base sets.
The edges of $\mathcal{G_{R,L}}$ are constructed based on the edges of $\mathcal{G_{R}}$: For each edge $(\mathcal{R}^{(i)}_{k}, \mathcal{R}^{(j)}_{k+1})$ in $\mathcal{G_{R}}$, all edges $(\mathcal{R}^{(i , v)}_{k}, \mathcal{R}^{(j , w)}_{k+1}) \, \forall \, v, w \in \{0, ...,m\}$ are added to $\mathcal{G_{R,L}}$.
Furthermore, we add weights to all edges according to the number of lane changes: $\mathcal{W}(\mathcal{R}^{(i , v)}_{k}, \mathcal{R}^{(i , w)}_{k+1}) = |w-v|$.
\begin{figure*}[t]
  \includegraphics[width=\textwidth]{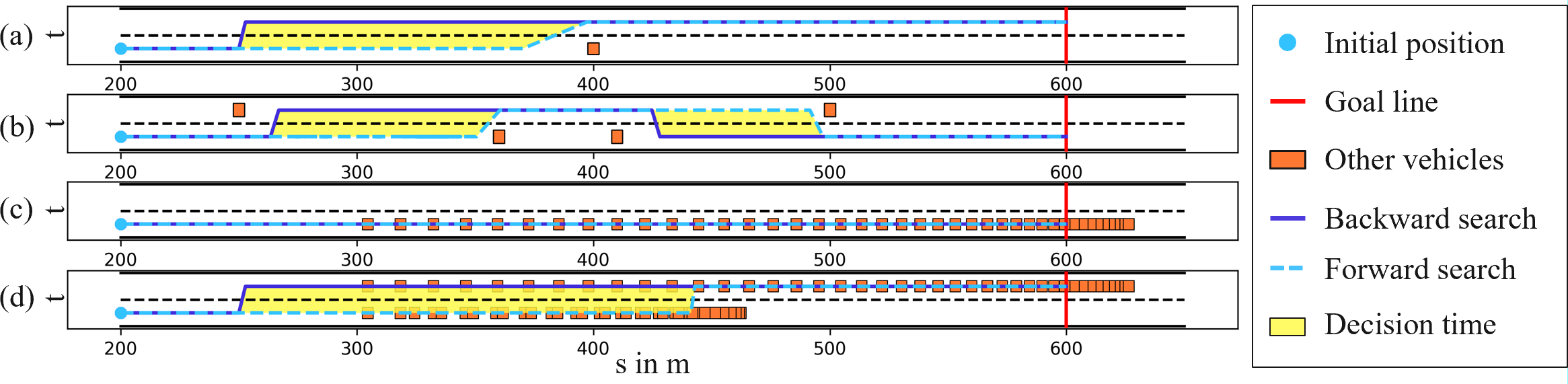}
  \caption{Visualization of the challenge description of the four evaluation scenarios. Each lane has a width of $3.75~m$ in t-direction. Only each fifth time step of the moving vehicles is plotted.}
  \label{fig:results}
\end{figure*}

\subsection{Graph Search and Filtering}

From the lane-aware reachability graph $\mathcal{G_{R,L}}$, the paths with the least lane changes from the initial position to the goal are determined by two graph searches.
The graph searches are conducted using the Dijkstra-algorithm.
First, a \emph{forward search} is conducted on the $\mathcal{G_{R,L}}$ to find all paths from the initial base set node to all base sets that are in the goal region while minimizing the accumulated edge weights.
Second, a \emph{backward search} is conducted on the reversed $\mathcal{G_{R,L}}$ starting from all goal region nodes and finding the cheapest paths to the initial position node.
Both searches provide a list of paths representing possible tactical maneuvers and from both the path which involves the least amount of lane changes is selected.
By the design of the Dijkstra-algorithm, the forward search finds tactical maneuvers with the latest lane change and the backwards search the ones with the earliest possible lane change.
These two paths are then combined to determine the decision time for the lane change(s).
The challenge description of the scenario consists of the number of the required lane changes of the easiest tactical maneuver as well as the corresponding decision times which represent the difficulty.

\section{EVALUATION} \label{sec:evaluation}

\subsection{Scenarios} \label{subsec:scenariogen}

We evaluate our method in four different scenarios that highlight different capabilities of the Challenge Description Method.
Each of the scenarios takes place on a straight two-lane highway with a lane width of 3.75~m.
The initial SUT position is on the right lane at $s=200~m$ with an initial velocity of 100 km/h which is a typical velocity for the rightmost lane on German highways.
The goal line is located at $s=600~m$.
The time between the discrete time steps is $0.1~s$.
Figure~\ref{fig:results} shows images of the scenarios together with a visualization of the challenge description, which is explained in Section~\ref{sec:results}.
In the scenarios (c) and (d), only each fifth time step of the moving vehicles is visualized.
The scenarios are characterized as follows:
\begin{itemize}
    \item (a) In this scenario, there is only one static vehicle in the right lane ahead of the initial SUT position.
    \item (b) In this scenario, there are four static vehicles. Two are blocking the SUT lane at about $s=375~m$ and $s=410~m$. Two more vehicles are located in the left lane before and after the two blocking vehicles at about $s=100~m$ and $s=500~m$.
    \item (c) The scenario features a vehicle that is starting in the SUT's lane at a THW of $3.6~s$ ahead of the SUT's initial position. The vehicle has the same starting velocity as the SUT. After a delay of $1~s$, it starts to decelerate with a constant deceleration of $1.3~\frac{m}{s^2}$.
    \item (d) This scenario features two decelerating lead vehicles. The vehicles start in the two lanes at a THW of $3.6~s$ ahead of the SUT's initial position. After a delay of $1~s$ the right vehicle starts to harshly decelerate with a constant deceleration of $3~\frac{m}{s^2}$. The left vehicle shows the same behavior as the vehicle in scenario~(c).
\end{itemize}

\subsection{Challenge Description} \label{sec:results}

To execute the Challenge Description Method on the evaluation scenarios, the eight parameters defining the normal operation of the ADS must be set (see Section~\ref{subsec:design}).
For the evaluation in this paper, general parameters that represent usual driving on a German highway were chosen.
The parameter values are given in Table~\ref{tab:parameters}

\begin{table}[t]
\caption{The eight parameters of the Challenge Description Method and their values in the evaluation.}
\begin{center}
\begin{tabular}{|c|c|c|}
\hline
\textbf{Symbol} & \textbf{Name} & \textbf{Value}\\
\hline
$v_{s,min}$ & Min. long. velocity & $60~km/h$ \\
$v_{s,max}$ & Max. long. velocity & $130~km/h$\\
$v_{t,min}$ & Min. lat. velocity & $-2~m/s$\\
$v_{t,max}$ & Max. lat. velocity & $2~m/s$\\
$a_{s,min}$ & Min. long. acceleration & $-4~m/s^2$\\
$a_{s,max}$ & Max. long. acceleration & $4~m/s^2$\\
$a_{t,min}$ & Min. lat. acceleration & $-2~m/s^2$\\
$a_{t,max}$ & Max. lat. acceleration & $2~m/s^2$\\
\hline
\end{tabular}
\label{tab:parameters}
\end{center}
\end{table}

Figure~\ref{fig:results} shows the four evaluation scenarios (a)\,-\,(d) together with a visualization of their respective challenge description.
For the visualization, first, the Challenge Description Method has been applied to the scenarios and, second, the resulting paths of the forward and backward search in the lane-aware reachability graph (see Section~\ref{subsec:larg}) have been expressed with a reference trajectory each.
This reference trajectory consists of one point per node in the path.
Each node corresponds to a lane-aware base set of the reachable set.
The s-coordinate of each point in the reference trajectory is the center of the base set in s-direction and the t-coordinate is the center of the lane the base set belongs to.
In Figure~\ref{fig:results}, the reference trajectories for the forward and backward search are shown with a solid blue and dashed light blue line, respectively.
Since the points of the reference trajectories connect base sets, lane changes happen in only one time step. Therefore, the reference trajectories do not represent a drivable trajectory, but they are the upper limit for what is theoretically possible to drive.

Table~\ref{tab:results} provides the results of the Challenge Description Method for the evaluation scenarios in the form of the minimum number of required lane changes and the decision time per lane change.
To successfully complete scenario~(a), at least one lane change is required and the decision time for this lane change is $6.0~s$.
Figure~\ref{fig:results} provides further details.
The forward search resulted in the latest possible lane change timing closely before the obstacle.
The backward search resulted in the earliest possible timing for a lane change at about $s=250~m$.
This timing marks the first point in time, the SUT could have completed the lane change and be entirely in the left lane.
For this lane change, the SUT would have to start moving laterally already at its starting point.
The yellow area that is enclosed by the two reference trajectories corresponds to the decision time for the lane change. 

Comparing scenarios (a) and (b), it can be seen that the additional vehicles in (b) were automatically taken into account by the Challenge Description Method.
The first vehicle on the left lane prevents an early lane change and the vehicle in the right lane is closer to the SUT's starting point than in (a).
Therefore, the decision time for the first required lane change is only $4.8~s$.
Because of the second vehicle in the left lane, a second lane change is required to reach the goal.
The decision time for this lane change is only $3.2~s$.
Any possible but not required lane changes are neglected, e.g., between the two vehicles on the right lane.

Scenarios (c) and (d) feature moving vehicles.
In scenario~(c), there is a slightly decelerating lead vehicle in front of the SUT's initial position.
Objectively, it is not possible to successfully complete the scenario without any action, i.\,e., deceleration and/or a lane change.
The Challenge Description Method determined that the required deceleration when staying in lane lays inside the boundaries for normal operation and, therefore, no lane change is required.
In scenario~(d), initially, the entire road is blocked by the two vehicles.
In contrast to (c), only decelerating in lane is not enough to successfully complete the scenario, since the vehicle in front comes to a standstill at approximately $s=470~m$.
When staying in lane, the SUT would also have to brake to a standstill and, thus, leave normal operation.
To stay in normal operation, it has to perform a lane change.
Since the left vehicle behaves the same as the vehicle in (c), the SUT can stay behind it and a second lane change is not required.

\begin{table}[tbp]
\caption{The challenge description of the four evaluation scenarios in the form of the number of required lane changes (LCs) and their respective decision times.}
\begin{center}
\begin{tabular}{|c|c|c|}
\hline
\textbf{Scenario} & \textbf{Required LCs} & \textbf{Decision time per LC}\\
\hline
(a)& 1 & $6.0~s$\\
(b)& 2 & $4.8~s$, $3.2~s$\\
(c)& 0 & -\\
(d)& 1 & $11.8~s$\\
\hline
\end{tabular}
\label{tab:results}
\end{center}
\end{table}

\subsection{Discussion and Outlook}

The results of Section~\ref{sec:results} showed that the Challenge Description Method is capable of objectively describing the challenge in different scenarios with various amounts of vehicles and their movements.
In contrast to previous methods, this method automatically considers the combined information of all scenes and there is no need to aggregate scene-level metric values to the scenario level.
The method does also not require complicated calibration, but only has eight easy-to-understand parameters.
In addition, the resulting challenge descriptions offer deeper and human-interpretable insights into the scenarios than previous methods.
Because the method analyzes the scenarios themselves without relying on detailed knowledge of any specific ADS, the resulting descriptions are also representative for multiple similar ADS.
Thus, these descriptions make it easier for ADS developers to select relevant test scenarios for their ADS.

Future work will focus on further detailing the tactical challenge by analyzing also the required speed adaptations in lane and on analyzing scenarios that do not have any possible tactical maneuver.

\section{CONCLUSION} \label{sec:conclusion}

This paper presented the Challenge Description Method that objectively describes the challenge of highway scenarios.
To make sure that all theoretically possible maneuvers of the SUT are considered, the method is based on a reachability analysis.
Instead of only a metric value per scenario, the method provides a human-interpretable description of the objective challenge for staying in normal operation.
Since the method does not depend on implementation details of any ADS, it can be applied to a scenario database in an offline pre-processing step and add additional human-interpretable information to the scenarios. Therefore, it is useful for selecting relevant test scenarios for multiple similar ADSs.







\bibliographystyle{IEEEtran}
\bibliography{IEEEabrv,references}

\end{document}